\documentstyle[12pt]{article}

\newcommand{\beq}{\begin{equation}}
\newcommand{\eeq}{\end{equation}}
\newcommand{\beqa}{\begin{eqnarray}}
\newcommand{\eeqa}{\end{eqnarray}}
\newcommand{\ba}{\begin{array}}
\newcommand{\ea}{\end{array}}

\begin{document}

\begin{flushright}
Preprint CAMTP/96-1\\
January 1996\\
\end{flushright}

\vskip 0.5 truecm

\begin{center}
\large
{\bf Quantum corrections to the semiclassical \\
quantization of a nonintegrable system}\\
\vspace{0.25in}
\normalsize
Luca Salasnich$^{(*)(+)}$\footnote{e--mail: luca.salasnich@uni-mb.si} 
and Marko Robnik$^{(*)}$\footnote{e--mail: robnik@uni-mb.si} \\
\vspace{0.2in}
$^{(*)}$ Center for Applied Mathematics and Theoretical Physics,\\
University of Maribor, Krekova 2, SLO-62000 Maribor, Slovenia\\
\vspace{0.2in}
$^{(+)}$ Dipartimento di Fisica "G. Galilei" della Universit\`a di Padova, \\
Istituto Nazionale di Fisica Nucleare, Sezione di Padova,\\
Via Marzolo 8, I--35131 Padova, Italy
\end{center}

\vspace{0.3in}

\normalsize
{\bf Abstract.} We study the semiclassical behaviour 
of a two--dimensional nonintegrable system. In particular we analyze 
the question of quantum corrections to the semiclassical quantization 
obtaining up to the second order of perturbation theory an explicit 
analytical formula for the energy levels, 
which is the usual semiclassical one plus quantum corrections. 
We compare the "exact" levels obtained numerically to the 
semiclassical levels studying also the effects of quantum corrections. 

\vspace{0.6in}

PACS numbers: 03.65.-w, 03.65.Ge, 03.65.Sq, 05.45.+b 

\vspace{0.6in}

Submitted to {\bf Journal of Physics A: Mathematical and General}
\normalsize
\vspace{0.1in}
  
\newpage

\par
Nowadays there is considerable renewed interest in the 
transition from classical mechanics to quantum mechanics, 
a powerful motivation behind that being the problem of the so--called 
quantum chaos (see Ozorio de Almeida 1990, Gutzwiller 1990, Casati and 
Chirikov 1995). 
An important aspect is represented by the semiclassical quantization 
formula of the (regular) energy levels for quasi--integrable systems 
(Maslov and Fredoriuk 1981, Rau 1992, Braun 1993), the so--called torus 
quantization, initiated by Einstein (1917) and completed by Maslov (1972). 
\par
It has been recently shown (Graffi and Paul 1987, Degli Esposti, Graffi and 
Herczynski 1991) that, for perturbed non--resonant harmonic oscillators, 
the algorithm of classical perturbation theory 
can be used to formulate the quantum mechanical perturbation theory 
as the semiclassically quantized classical perturbation theory equipped 
with the quantum corrections in powers of $\hbar$ "correcting" 
the classical Hamiltonian that appears in the classical algorithm. 
In effect, one can explicitly calculate corrections to the 
Einstein--Brillouin--Keller (EBK) quantization of the classical tori 
(Maslov and Fredouriuk 1981). 
For example, the quantum corrections of the one--dimensional $x^4$ perturbed 
harmonic oscillator have been studied in great details by Alvarez, Graffi 
and Silverstone (1988). 
\par
Examples of rather detailed studies of semiclassical approximations, 
their resummations and of the Birkkoff--Gustavson normal forms can be found 
in Ali, Wood and Devitt (1986) and Ali and Devitt (1989). 
Another example of a rather complete 
semiclassical analysis of a one--dimensional system, namely 
the quartic oscillator, has been published by Voros (1983).
\par
The aim of this paper is to extend previous studies 
to a two--dimensional system, which is more interesting because 
it is nonintegrable and thus generic. 
The integrable systems are rather exceptional in the 
sense that they are typically isolated points in the functional space of 
Hamiltonians and their measure is zero in this space. 
If we randomly choose a system realized in nature, the probability is one 
that the system is nonintegrable (see Robnik 1995). 
\par
The model is given by two non--resonant oscillators 
coupled by a nonlinear quartic interaction of strength $g$ (Pullen and 
Edmonds 1981): 
\beq
H={\omega_1 \over 2}(p_1^2+q_1^2)+{\omega_2 \over 2}(p_2^2+q_2^2)
+ g q_1^2q_2^2. 
\eeq
Note that a similar Hamiltonian has been obtained for the 
Yang--Mills--Higgs quantum mechanics (Savvidy 1984, Salasnich 1995). 
\par
Through the canonical transformation in 
action--angle variables (see Born 1927 and Dittrich and Reuter 1992): 
\beq
q_k=\sqrt{2I_k}\cos{\theta_k}, \;\;\; p_k=\sqrt{2I_k}\sin{\theta_k}, 
\;\;\; k=1,2 
\eeq
the Hamiltonian can be written: 
\beq
H=H_0(I_1,I_2)+ g V(I_1,I_2,\theta_1,\theta_2),
\eeq
where: 
\beq
H_0(I_1,I_2)=\omega_1 I_1+ \omega_2 I_2,
\eeq
\beq
V(I_1,I_2,\theta_1,\theta_2)=4I_1I_2\cos^2{\theta_1}\cos^2{\theta_2}.
\eeq
\par
Following the classical perturbation theory (see Born 1927, Dittrich 
and Reuter 1992), we search for a canonical transformation 
$(I_1,I_2,\theta_1,\theta_2) \to ({\tilde I}_1,{\tilde I}_2,
{\tilde \theta}_1,{\tilde \theta}_2)$ to obtain a new Hamiltonian that 
depends only on the new action variables up to the second order in a 
power series of $g$:
\beq
{\tilde H}({\tilde I}_1,{\tilde I}_2)=
{\tilde H}_0({\tilde I}_1,{\tilde I}_2)
+g {\tilde H}_1({\tilde I}_1,{\tilde I}_2)
+g^2 {\tilde H}_2({\tilde I}_1,{\tilde I}_2).
\eeq
\par
The generator $S$ of the canonical transformation is supposed to be capable 
of expansion as a power series in $g$ of the form:
\beq
S({\tilde I}_1,{\tilde I}_2,\theta_1,\theta_2)=
{\tilde I}_1\theta_1 + {\tilde I}_2\theta_2 
+g S_1({\tilde I}_1,{\tilde I}_2,\theta_1,\theta_2)
+g^2 S_2({\tilde I}_1,{\tilde I}_2,\theta_1,\theta_2), 
\eeq
and to satisfy the equations:
\beq
I_k={\partial S\over \partial \theta_k}={\tilde I}_k
+g {\partial S_1\over \partial \theta_k}
+g^2 {\partial S_2\over \partial \theta_k},
\eeq
\beq
{\tilde \theta}_k={\partial S\over \partial {\tilde I}_k}=\theta_k
+g {\partial S_1\over \partial {\tilde I}_k}
+g^2 {\partial S_2\over \partial {\tilde I}_k}. 
\eeq
From the Hamilton--Jacobi equation: 
\beq
H_0({\partial S\over \partial \theta_1},{\partial S\over \partial \theta_2})+
V({\partial S\over \partial \theta_1},{\partial S\over \partial 
\theta_2}, \theta_1, \theta_2)=
{\tilde H}_0({\tilde I}_1,{\tilde I}_2)
+g {\tilde H}_1({\tilde I}_1,{\tilde I}_2)
+g^2 {\tilde H}_2({\tilde I}_1,{\tilde I}_2),
\eeq
we have a number of differential equations that result on equating the 
coefficients of the powers of $g$:
\beq
{\tilde H}_0({\tilde I}_1,{\tilde I}_2)=
H_0({\tilde I}_1,{\tilde I}_2)=
\omega_1{\tilde I}_1+\omega_2{\tilde I}_2,
\eeq
\beq
{\tilde H}_1({\tilde I}_1,{\tilde I}_2)=
(\omega_1{\partial S_1\over \partial \theta_1} +
\omega_2{\partial S_1\over \partial \theta_2} )
+ V({\tilde I}_1,{\tilde I}_2, \theta_1, \theta_2),
\eeq
\beq
{\tilde H}_2({\tilde I}_1,{\tilde I}_2)=
( \omega_1{\partial S_2\over \partial \theta_1} 
+\omega_2 {\partial S_2\over \partial \theta_2} )
+ ( {\partial V\over \partial I_1}{\partial S_1\over \partial \theta_1} 
+{\partial V\over \partial I_2}{\partial S_1\over \partial \theta_2} ).
\eeq
The unknown functions ${\tilde H}_1$, $S_1$, ${\tilde H}_2$ and $S_2$ 
may be determined by averaging over the time variation of the 
unperturbed motion. \\
At the first order in $g$ we obtain: 
\beq
{\tilde H}_1({\tilde I}_1,{\tilde I}_2)= 
{1\over 4\pi^2}\int_0^{2\pi}\int_0^{2\pi} d\theta_1 d\theta_2 
V({\tilde I}_1,{\tilde I}_2, \theta_1, \theta_2) 
={\tilde I}_1 {\tilde I}_2 ,
\eeq
and 
$$
S_1({\tilde I}_1,{\tilde I}_2,\theta_1,\theta_2)=
-{1\over 4}{\tilde I}_1 {\tilde I}_2 
[{2\over \omega_1}\sin{2\theta_1}+{2\over \omega_2}\sin{2\theta_2}+
$$
\beq
+{1\over \omega_1 - \omega_2}\sin{2(\theta_1 -\theta_2 )}+
{1\over \omega_1 + \omega_2}\sin{2(\theta_1 +\theta_2 )}].
\eeq
At the second order in $g$ we have: 
$$
{\tilde H}_2({\tilde I}_1,{\tilde I}_2)=
{1\over 4\pi^2}\int_0^{2\pi}\int_0^{2\pi} d\theta_1 d\theta_2 
( {\partial V\over \partial I_1}{\partial S_1\over \partial \theta_1} 
+{\partial V\over \partial I_1}{\partial S_1\over \partial \theta_2} )
$$
\beq
=-{1\over 8}{\tilde I}_1{\tilde I}_2
[4 ({ {\tilde I}_1\over \omega_2 }+{ {\tilde I}_2\over \omega_1 })
- {({\tilde I}_1-{\tilde I}_2)\over \omega_1 -\omega_2}
+ {({\tilde I}_1+{\tilde I}_2)\over \omega_1 +\omega_2}].
\eeq
\par
The approximate Hamiltonian (6) depends only on the actions
\footnote{Note that the integrable approximate Hamiltonian (6) could be 
obtained alternatively as the Birkhoff--Gustavson normal form, which is 
a purely algebraic method of calculating the action variables order by 
order for perturbed harmonic oscillators with polynomial perturbations 
(Robnik 1984, Kalu\v za and Robnik 1992, Kalu\v za 1993, Robnik 1993).}, 
so that by an application of the EBK rule: 
\beq
{\tilde I}_{k}=(n_k+{1\over 2})\hbar, 
\eeq 
to equations (11), (14) and (16), we obtain a semiclassical 
analytical formula of the energy levels. 
\par
Now we show how to connect this semiclassical formula 
with the usual quantum perturbation theory. 
In quantum mechanics the generalized coordinates satisfy the usual 
commutation rules $[{\hat q}_k,{\hat p}_l]=i\hbar \delta_{kl}$, 
with $k,l=1,2$. Introducing the creation and destruction operators:
\beq
{\hat a}_k={1\over \sqrt{2\hbar}}({\hat q}_k+i {\hat p}_k),
\;\;\;\;
{\hat a}_k^+ ={1\over \sqrt{2\hbar}}({\hat q}_k -i {\hat p}_k),
\eeq
the quantum Hamiltonian can be written:
\beq
{\hat H}={\hat H}_0 + g {\hat V},
\eeq
where:
\beq
{\hat H}_0=\hbar \omega_1 ({\hat a}_1^+ {\hat a}_1 +{1\over 2})+
\hbar \omega_2 ({\hat a}_2^+ {\hat a}_2 +{1\over 2}),
\eeq
\beq
{\hat V}= {\hbar^2 \over 4} ({\hat a}_1 +{\hat a}_1^+)^2 
({\hat a}_2 +{\hat a}_2^+)^2. 
\eeq
\par
At this point we mention the general problem of quantization (Robnik 1984, 
Robnik 1988, Abraham and Marsden 1978): 
There is no unique quantization prescription. 
If one chooses certain quantization procedure, this will 
{\it not} commute in general with the classical canonical transformations. 
However, the quantization does commute with the {\it linear} classical 
canonical transformations. This is exactly the approach implemented in 
our present case (18), and thus our quantization is equivalent 
to the coordinate space quantization (the latter one yields the 
{\it right} quantum mechanics, whose results agree with experiments). 
\par
If $|n_1 n_2>$ is the basis of the occupation numbers of the two 
harmonic oscillators, the matrix elements are: 
\beq
<n_{1}^{'}n_{2}^{'}|{\hat H}_0|n_{1}n_{2}>=\hbar [\omega_1 (n_{1}+{1\over 2})+
\omega_2 (n_{1}+{1\over 2})]
\delta_{n_{1}^{'}n_{1}} \delta_{n_{2}n_{2}} ,
\eeq
and:
$$
<n_{1}^{'}n_{2}^{'}|{\hat V}|n_{1}n_{2}>=
{\hbar^2 \over 4}
[\sqrt{n_{1}(n_{1}-1)} \delta_{n^{'}_{1}n_{1}-2}
+\sqrt{(n_{1}+1)(n_{1}+2)}\delta_{n^{'}_{1}n_{1}+2}+
(2n_{1}+1)\delta_{n^{'}_{1}n_{1}}]\times 
$$
\beq
\times[\sqrt{n_2 (n_2-1)}\delta_{n^{'}_2 n_2-2}+ \sqrt{(n_2+1)(n_2+2)}
\delta_{n^{'}_2 n_2+2}+ (2n_2+1)\delta_{n^{'}_2 n_2}] .
\eeq
The Rayleigh--Schr\"odinger perturbation theory (see Messiah 1962) up to the 
second order gives us:
\beq
E(n_1\hbar ,n_2\hbar )=E_{0}(n_1\hbar ,n_2\hbar )+g E_{1}(n_1\hbar ,n_2\hbar )+
g^2 E_{2}(n_1\hbar ,n_2\hbar ) , 
\eeq
where: 
\beq
E_{0}(n_1\hbar ,n_2\hbar )=\hbar [\omega_1 (n_{1}+{1\over 2})+
\omega_2 (n_{1}+{1\over 2})],
\eeq
\beq
E_{1}(n_1\hbar ,n_2\hbar )=<n_{1}n_{2}|{\hat V}|n_{1}n_{2}>,
\eeq
\beq
E_{2}(n_1\hbar ,n_2\hbar )= 
\sum_{\stackrel{n_1^{'}n_2^{'}}{(n_1^{'},n_2^{'})\neq (n_1,n_2)}} 
{ |<n_{1}^{'}n_{2}^{'}|{\hat V}|n_{1}n_{2}>|^2 \over 
\hbar [\omega_1 (n_1 -n_1^{'}) + \omega_2 (n_2 -n_2^{'})] }.
\eeq
\par
We obtain immediately:
\beq
E_{1}(n_1\hbar ,n_2\hbar )=\hbar^2 
(n_1+{1\over 2})(n_2+{1\over 2}),
\eeq
and after some calculations:
$$
E_{2}(n_1\hbar ,n_2\hbar )= 
{\hbar^3\over 32}[
{n_1(n_1-1)n_2(n_2-1)\over \omega_1+\omega_2}-
{(n_1+1)(n_1+2)(n_2+1)(n_2+2)\over \omega_1+\omega_2}+
$$
$$
+{n_1(n_1-1)(n_2+1)(n_2+2)\over \omega_1-\omega_2}-
{(n_1+1)(n_1+2)n_2(n_2-1)\over \omega_1-\omega_2}+
$$
$$
+{n_1(n_1-1)(2n_2+1)^2\over \omega_1}-
{(n_1+1)(n_1+2)(2n_2+1)^2\over \omega_1}+
$$
\beq
+{(2n_1+1)^2n_2(n_2-1)\over \omega_2}-
{(2n_1+1)^2(n_2+1)(n_2+2)\over \omega_2}  ].
\eeq
The zero and first order quantum terms coincide with the semi-classical ones:
\beq
E_{0}(n_1\hbar ,n_2\hbar )={\tilde H}_{0}
((n_1+{1\over 2})\hbar ,(n_2+{1\over 2})\hbar ), 
\eeq
\beq
E_{1}(n_1\hbar ,n_2\hbar )={\tilde H}_{1}
((n_1+{1\over 2})\hbar ,(n_2+{1\over 2})\hbar ),
\eeq
and the second order quantum term can be written: 
\beq
E_{2}(n_1\hbar ,n_2\hbar )={\tilde H}_{2}
((n_1+{1\over 2})\hbar ,(n_2+{1\over 2})\hbar ) + \hbar^2 
Q_2((n_1+{1\over 2})\hbar ,(n_2+{1\over 2})\hbar ),
\eeq
where:
\beq
Q_2({\tilde I}_1,{\tilde I}_2)= - {3\over 32}
[{({\tilde I}_1 - {\tilde I}_2 )\over \omega_1 -\omega_2} + 
{({\tilde I}_1 + {\tilde I}_2 )\over \omega_1 +\omega_2}].
\eeq
The quantum series rearranges directly 
into the classical canonical perturbation series plus quantum corrections 
proportional to successive powers of $\hbar$ (see Degli 
Esposti, Graffi and Herczynski 1991). 
\par
The term $\hbar^2 Q_2$ represents the quantum corrections 
to the EBK quantization up to the second order of perturbation theory. 
These quantum corrections depend linearly 
on quantum numbers. It is in contrast with first order results in 
a previous paper of Robnik (1984) in which the spectra differ only by an 
additive constant independent of the quantum numbers (actions).  
\par
We compute the "exact" levels with a numerical diagonalization 
of the truncated matrix of the Hamiltonian (19) 
in the basis of the unperturbed oscillators (see Graffi, Manfredi and 
Salasnich 1994). 
The numerical energy levels depend on the dimension of the truncated matrix: 
We compute the numerical levels in double precision 
increasing the matrix dimension until the first 100 levels converge 
within $8$ digits (matrix dimension $1225\times 1225$). 
\par
Then we compare the "exact" levels to the levels 
of the quantum perturbation theory and to the semiclassical results. 
A very good agreement is observed for the 
lowest energy levels (see Table 1). 
\par
In table 2 we show the error in units of the mean level spacing $D$ 
between the "exact" levels and the levels obtained with semiclassical 
and quantum perturbation theory. 
\par
We observe that the algorithm provided 
by the appropriate semiclassical quantization is comparable to the algorithm 
provided by ordinary quantum perturbation theory but 
the quantal corrections do not always increase the accuracy. 
Thus for some of the calculated levels the semiclassical quantization 
gives better results than the quantum perturbation theory. 
In fact, Rayleigh--Schr\"odinger perturbation theory diverges for 
any value of $g$, but it has been proved to be Borel summable 
to the exact energy levels only for some special systems 
like the anharmonic oscillators with $f$ degrees of freedom 
with a polynomial perturbing potential which is 
asymptotically positive definite (Graffi, Grecchi and Simon 1970, 
Reed and Simon 1978, Caliceti, Graffi and Maioli 1980). 
\par
In Figure 1 we plot the error, in units of the mean level spacing $D$, 
between the "exact" levels and the semiclassical levels. The figure shows some 
systematics but the prediction of individual levels by EBK quantization 
worsens as the quantum number increases, contrary to the naive expectation 
(see also Prosen and Robnik 1993). 
As is seen, by decreasing $\hbar$ from $1$ to $10^{-1}$ the quality of the 
approximation improves considerably but only for sufficiently low levels 
(see also Table 3). 
\par
If $\hbar$, no matter how small, is kept fixed, the EBK quantization 
(torus quantization) of the individual levels is only a first order 
approximation of an expansion in $\hbar$. 
Therefore, the accuracy of the approximation decreases for higher levels. 
To get a good agreement it is necessary, as is well known, to implement 
the classical limit, i.e. $\hbar \to 0$ and $n_1,n_2 \to \infty$, 
while at the same time keeping the actions 
${\tilde I}_1 =(n_1 +1/2)\hbar$ and ${\tilde I}_2 =(n_2 +1/2)\hbar$ 
constant (Graffi, Manfredi and Salasnich 1994). 
\par
In Figure 2 we plot the error, in units of the mean level spacing $D$, 
between semiclassical levels and levels obtained 
with quantum perturbation theory. Also in this case the error 
increases for higher levels. 
\par
In conclusion, we have examined the transition between the classical 
and the quantum mechanics for a two--dimensional, nonintegrable 
and non--resonant system. 
Up to the second order of perturbation theory 
we have decomposed the quantum description into 
the classical description (i.e. the leading semiclassical term) 
plus quantum corrections which depend linearly on the quantum numbers. 
The analytical energy levels are in good agreement 
with the "exact" numerical ones, 
the semiclassical quantization is comparable to the 
quantum perturbation theory, and for some levels the semiclassical 
quantization gives better results than quantum perturbation theory.  
\par 
Finally we note that the extraction of quantum corrections for resonant systems 
is a more intricate procedure; some initial results for perturbed 
resonant oscillators can be found in Graffi (1992).
\par
The present paper supports the following {\it general conclusions}. 
\par
(c1) The quantum corrections, defined as the difference between the 
semiclassical and the quantum perturbation terms of the same order, 
as derived by Degli Esposti, Graffi and Herczynski (1991), certainly 
vanish in the limit $\hbar \to 0$.
\par
(c2) The above conclusion, however, changes drastically when the quantum 
corrections are measured in natural units of the mean level spacing $D$, 
which scales as $D\sim \hbar^{f}$, where $f$ is the number of freedoms: 
The quantum corrections and their low $\hbar$--power terms can even diverge 
or remain constant. 
This is implicit in Degli Esposti, Graffi and Herczynski (1991). 
Probably the same conclusion applies when exact levels are compared to the 
semiclassical ones, supported by the results in Prosen and Robnik (1993) 
and by the results of the present paper.
\par
(c3) Since both the classical and quantum perturbation series typically 
diverge and thus do not necessarily describe the exact levels, 
not even after a certain resummation (except for some important notable 
exceptions previously described (Graffi, Grecchi and Simon 1970)), 
it is important to compare the 
semiclassical approximation (and the quantal perturbation results) 
with the exact spectra, which in general is impossible, since we 
generally do not have explicit solutions of the Schr\"odinger problem 
in a closed form. Therefore, we stress the importance of specific 
case studies like the present one, in order to get a better understanding 
of the quality of semiclassical approximations.

\begin{center}
{\bf Acknowledgements}
\end{center}

\par
LS thanks Professors Gabriel Alvarez and Sandro Graffi 
for many enlightening discussions. 

\newpage

\parindent=0.pt

\section*{References} 
\vspace{0.6 cm}

Ali M K and Wood W R 1989 {\it J. Math. Phys.} {\bf 30} 1238 
\\\\
Ali M K, Wood W R and Devitt J S 1986 {\it J. Math. Phys.} {\bf 27} 1806 
\\\\
Alvarez G, Graffi S and Silverstone H J 1988 {\it Phys. Rev.} A {\bf 38} 1687 
\\\\
Braun P A 1993 {\it Rev. Mod. Phys.} {\bf 65} 115 
\\\\
Born M 1927 {\it The Mechanics of the Atom} 
(London: G. Bell and Sons Ltd.) 
\\\\
Caliceti E, Graffi S and Maioli M 1980 {\it Commun. Math. Phys.} 
{\bf 75} 51 
\\\\
Casati G and Chirikov B 1995 {\it Quantum Chaos} (Cambridge: Cambridge 
University Press) 
\\\\
Degli Esposti M, Graffi S and Herczynski J 1991 {\it Ann. Phys.} (N.Y.) 
{\bf 209} 364 
\\\\
Dittrich W and Reuter M 1992 {\it Classical and Quantum Dynamics} 
(New York: Springer) 
\\\\
Einstein A 1917 {\it Verhandlungen der Deutschen Physikalischen 
Gefsellschaft} {\bf 19} 82 
\\\\
Messiah A 1962 {\it Mecanique Quantique} (Paris: Dunod) 
\\\\ 
Graffi S 1992 {\it Probabilistic Methods in Mathematical Physics}, 
Ed. by Guerra F, Loffredo M I and Marchioro C 
(Singapore: World Scientific) 
\\\\
Graffi S, Grecchi V and Simon B 1970 {\it Phys. Lett.} B {\bf 32} 631 
\\\\
Graffi S, Manfredi V R and Salasnich L 1994 
{\it Nuovo Cim.} B {\bf 109} 1147 
\\\\
Graffi S and Paul T 1987 {\it Commu. Math. Phys.} {\bf 107} 25 
\\\\
Gutzwiller M C 1990 {\it Chaos in Classical and Quantum Mechanics} 
(New York: Springer) 
\\\\
Kalu\v za M 1993 {\it Comp. Phys. Commun.} {\bf 74} 441
\\\\
Kalu\v za M and Robnik M 1992 {\it J. Phys.} A {\bf 25} 5311
\\\\
Maslov V P 1972 {\it Theorie des Perturbations et Methodes Asymptotiques} 
(Paris: Dunod) (1965 Russian edition) 
\\\\
Maslov V P and Fredoriuk M V 1981 {\it Semiclassical Approximation in 
Quantum Mechanics} (London: Reidel Publishing Company) 
\\\\
Ozorio de Almeida A M 1990 {\it Hamiltonian Systems: Chaos and 
Quantization} (Cambridge: Cambridge University Press) 
\\\\
Pullen R A and Edmonds R A 1981 {\it J. Phys.} A {\bf 14} L477 
\\\\
Prosen T and Robnik M 1993 {\it J. Phys.} A {\bf 26} L37 
\\\\
Rau A R P 1992 {\it Rev. Mod. Phys.} {\bf 64} 623 
\\\\
Reed M and Simon B 1978 {\it Methods of Modern Mathematical Physics} 
(New York: Academic Press) 
\\\\
Robnik M 1984 {\it J. Phys.} A {\bf 17} 109 
\\\\
Robnik M 1988 {\it Lecture Notes Max--Planck--Institute for Kernphysik}; 
in preparation for Lecture Notes in Physics (New York: Springer) 
\\\\
Robnik M 1993 {\it J. Phys.} A {\bf 26} 7427 
\\\\
Robnik M 1995 to appear in {\it Proceedings of the Summer School 
"Complexity and Chaotic Dynamics of Nonlinear Systems"}, Xanthi, Greece, 
on 17--28 July 1995. 
\\\\
Savvidy G K 1984 {\it Nucl. Phys.} B {\bf 246} 302 
\\\\
Salasnich L 1995 {\it Phys. Rev.} D {\bf 52} 6189 
\\\\
Voros A 1983 {\it Ann. Inst. H. Poincar\`e} A {\bf 39} 211 

\newpage

\begin{center}
\begin{tabular}{|ccc|} \hline\hline 
  $E^{ex}$ & $E^{sc}$ & $E^{qp}$  \\ \hline
  1.230722 & 1.230990 & 1.230522  \\
  2.275974 & 2.273214 & 2.274701  \\
  2.689415 & 2.690856 & 2.687816  \\
  3.316524 & 3.308447 & 3.311808  \\
  3.820434 & 3.814018 & 3.812833  \\
  4.146646 & 4.148302 & 4.142610  \\
  4.354307 & 4.336609 & 4.341846  \\
  4.937708 & 4.915967 & 4.916677  \\
  5.359848 & 5.347322 & 5.345305  \\
  5.390110 & 5.357700 & 5.364811  \\
  5.603778 & 5.603248 & 5.594904  \\   
  6.047742 & 5.996702 & 5.999287  \\ 
  6.424398 & 6.371719 & 6.380706  \\ 
  6.546966 & 6.510986 & 6.509044  \\ 
  6.897049 & 6.873125 & 6.866657  \\ 
  7.062932 & 7.055694 & 7.044699  \\ 
  7.152476 & 7.056224 & 7.060684  \\ 
  7.457506 & 7.378668 & 7.389530  \\ 
  7.723943 & 7.639295 & 7.639228  \\ 
  8.144146 & 8.093505 & 8.088912  \\ 
  8.253060 & 8.094533 & 8.100868  \\ 
  8.435119 & 8.378546 & 8.382309  \\ 
  8.489923 & 8.391429 & 8.391282  \\ 
  8.528618 & 8.505639 & 8.491993  \\ 
  8.892640 & 8.732248 & 8.734056  \\
\hline\hline
\end{tabular}
\end{center}

\vskip 0.5 truecm

{\bf Table 1}: Comparison between "exact" levels 
and levels obtained by perturbation theories. First 25 levels. 
$E^{ex}$ are "exact" levels, $E^{sc}$ are semiclassical levels,  
and $E^{qp}$ are levels obtained with quantum perturbation theory.  
$\hbar =1$, $g=10^{-1}$, $\omega_1=1$ and $\omega_2=\sqrt{2}$. 

\newpage 

\begin{center}
\begin{tabular}{|cc|} \hline\hline 
     $|E^{ex}-E^{sc}|/D$   &     $|E^{ex}-E^{qp}|/D$   \\ \hline
  1.0611359$\cdot 10^{-3}$ &  1.1284242$\cdot 10^{-3}$ \\
  1.5578579$\cdot 10^{-2}$ &  7.1859419$\cdot 10^{-3}$ \\
  8.1338054$\cdot 10^{-3}$ &  9.0260478$\cdot 10^{-3}$ \\
  4.5591835$\cdot 10^{-2}$ &  2.6613854$\cdot 10^{-2}$ \\
  3.6215890$\cdot 10^{-2}$ &  4.2791300$\cdot 10^{-2}$ \\
  9.3476856$\cdot 10^{-3}$ &  2.2781115$\cdot 10^{-2}$ \\
  9.9898852$\cdot 10^{-2}$ &  7.0337765$\cdot 10^{-2}$ \\
  0.1227176                &  0.1187100                \\
  7.0703819$\cdot 10^{-2}$ &  9.2249520$\cdot 10^{-2}$ \\
  0.1829406                &  0.1428019                \\
  2.9902905$\cdot 10^{-2}$ &  5.0089385$\cdot 10^{-2}$ \\
  0.2880960                &  0.2735053                \\
  0.2973495                &  0.2466222                \\
  0.2030921                &  0.2140520                \\
  0.1350395                &  0.1715501                \\
  4.0854741$\cdot 10^{-2}$ &  0.1029161                \\
  0.5432989                &  0.5181223                \\
  0.4450069                &  0.3836938                \\
  0.4778005                &  0.4781800                \\
  0.2858459                &  0.3117708                \\
  0.8948155                &  0.8590558                \\
  0.3193286                &  0.2980870                \\
  0.5559518                &  0.5567808                \\
  0.1297049                &  0.2067311                \\
  0.9053394                &  0.8951331                \\
\hline\hline
\end{tabular}
\end{center}

\vskip 0.5 truecm

{\bf Table 2}: The error measured in units 
of the mean level spacing $D$ for the first 25 levels. 
$D$ is calculated for the lowest $100$ levels. 
$E^{ex}$ are "exact" levels, $E^{sc}$ are semiclassical levels,  
and $E^{qp}$ are levels obtained with quantum perturbation theory.  
$\hbar =1$, $g=10^{-1}$, $\omega_1=1$ and $\omega_2=\sqrt{2}$. 

\newpage

\begin{center}
\begin{tabular}{|cc|} \hline\hline 
                       $\hbar= 1$   &     $\hbar =10^{-1}$      \\ \hline
           1.0611359$\cdot 10^{-3}$ &  2.4773894$\cdot 10^{-5}$ \\
           1.5578579$\cdot 10^{-2}$ &  1.0003044$\cdot 10^{-4}$ \\
           8.1338054$\cdot 10^{-3}$ &  1.8136360$\cdot 10^{-4}$ \\
           4.5591835$\cdot 10^{-2}$ &  2.5054353$\cdot 10^{-4}$ \\
           3.6215890$\cdot 10^{-2}$ &  5.6091835$\cdot 10^{-6}$ \\
           9.3476856$\cdot 10^{-3}$ &  3.1972348$\cdot 10^{-4}$ \\
           9.9898852$\cdot 10^{-2}$ &  4.3938606$\cdot 10^{-4}$ \\
           0.1227176                &  2.3371598$\cdot 10^{-4}$ \\
           7.0703819$\cdot 10^{-2}$ &  9.3486393$\cdot 10^{-5}$ \\
           0.1829406                &  7.0301769$\cdot 10^{-4}$ \\
           2.9902905$\cdot 10^{-3}$ &  4.4125578$\cdot 10^{-4}$ \\
           0.2880960                &  6.3570746$\cdot 10^{-4}$ \\
           0.2973495                &  2.3932516$\cdot 10^{-4}$ \\
           0.2030921                &  1.0582660$\cdot 10^{-3}$ \\
           0.1350395                &  1.1966258$\cdot 10^{-4}$ \\
           4.0854741$\cdot 10^{-2}$ &  1.2452388$\cdot 10^{-3}$ \\
           0.5432989                &  5.2726327$\cdot 10^{-4}$ \\
           0.4450069                &  8.1146188$\cdot 10^{-4}$ \\
           0.4778005                &  1.5294374$\cdot 10^{-3}$ \\
           0.2858459                &  3.0663537$\cdot 10^{-4}$ \\
           0.8948155                &  2.1427081$\cdot 10^{-3}$ \\
           0.3193286                &  5.2352381$\cdot 10^{-5}$ \\
           0.5559518                &  1.7238891$\cdot 10^{-4}$ \\
           0.1297049                &  2.1165321$\cdot 10^{-3}$ \\
           0.9053394                &  5.6091836$\cdot 10^{-4}$ \\
\hline\hline
\end{tabular}
\end{center}

\vskip 0.5 truecm

{\bf Table 3}: The error measured in units 
of the mean level spacing $D$ between "exact" levels and 
semiclassical levels. First 25 levels. 
$D$ is calculated for the lowest $100$ levels. 
$g=10^{-1}$, $\omega_1=1$ and $\omega_2=\sqrt{2}$. 

\newpage

\parindent=0.pt

\section*{Figure Captions} 
\vspace{0.6 cm}

{\bf Figure 1:} The error measured in units of the mean level spacing 
between "exact" levels and semiclassical levels. First 100 levels. 
$g= 10^{-1}$, $\omega_1=1$ and $\omega_2=\sqrt{2}$. 
(a) $\hbar =1$, (b) $\hbar =10^{-1}$. 

\vskip 0.5 truecm

{\bf Figure 2:} The error measured in units of the mean level spacing 
between semiclassical levels and levels obtained with quantum 
perturbation theory. First 100 levels. 
$g= 10^{-1}$, $\omega_1=1$ and $\omega_2=\sqrt{2}$. 
(a) $\hbar =1$, (b) $\hbar =10^{-1}$. 

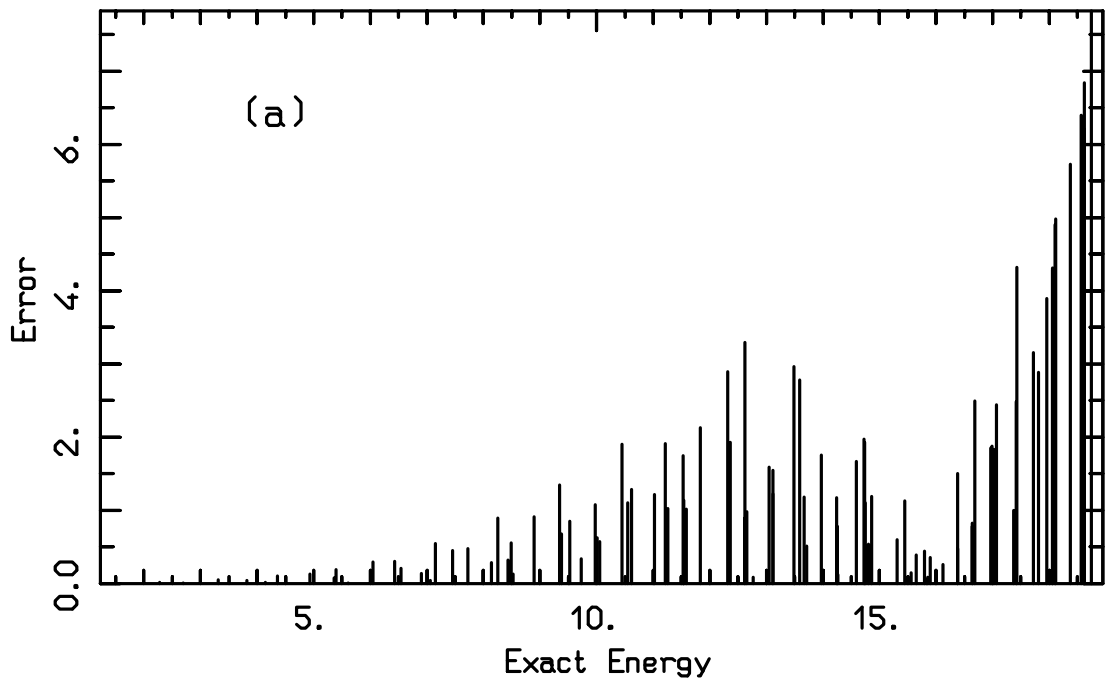
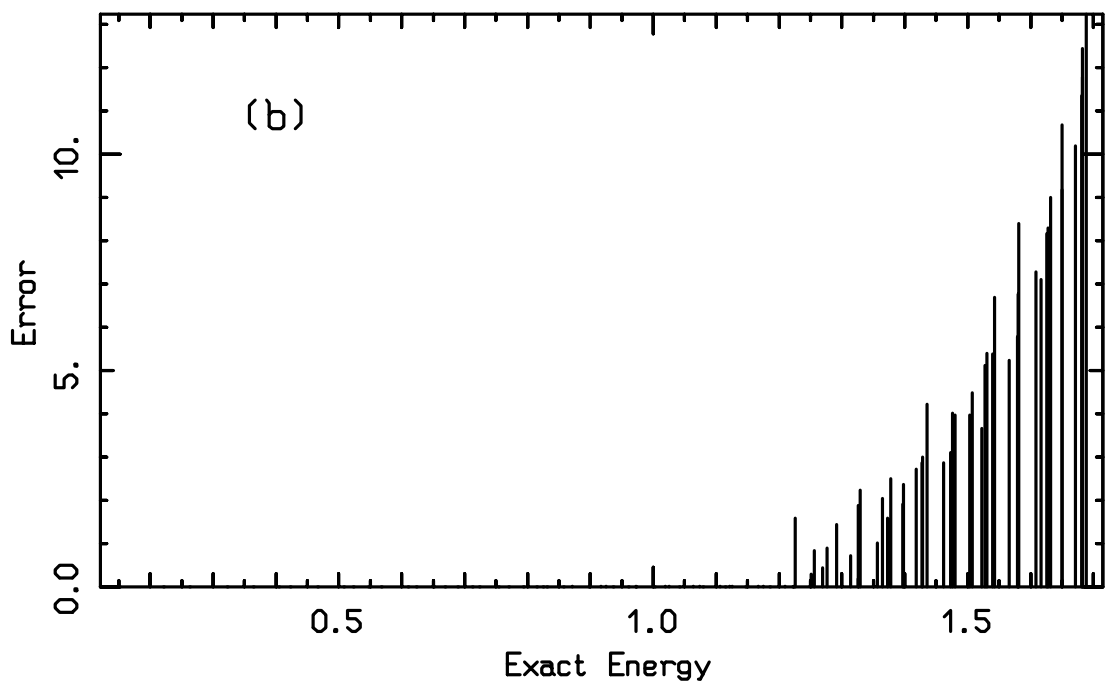
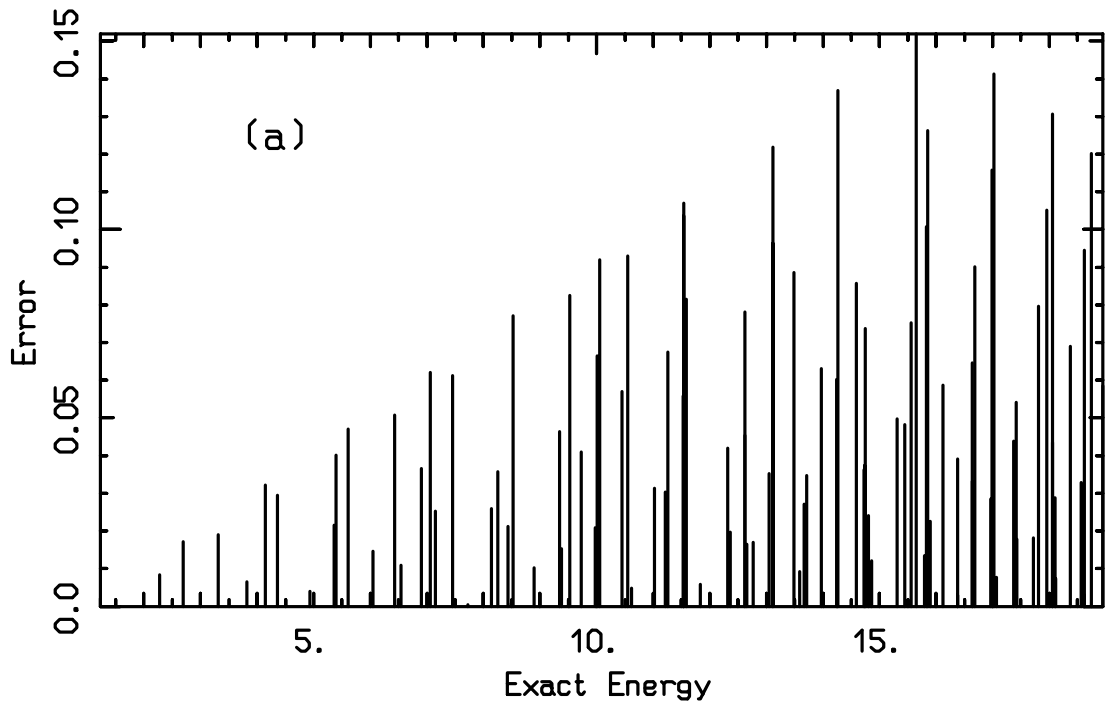
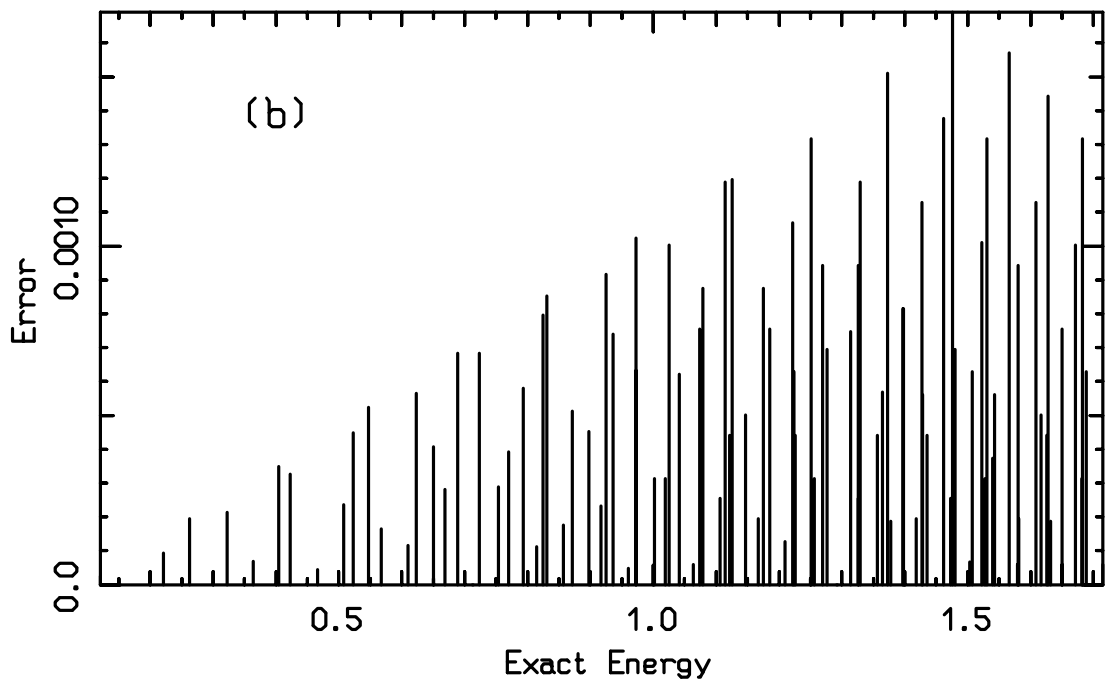
\end{document}